\documentclass[cits,hyper]{PoS}
\usepackage{amsmath}
\usepackage{overpic}
\usepackage{axodraw}

\title{Chiral Perturbation Theory and Mesons}

\ShortTitle{Chiral Perturbation Theory and Mesons}

\author{\speaker{Johan Bijnens}\\
        Department of Astronomy and Theoretical Physics, Lund University\\
        S\"olvegatan 14A, SE 22362 Lund, Sweden\\
        E-mail: \email{bijnens@thep.lu.se}}

\dedicated{dedicated to Ximo Prades 1963-2010}

\abstract{The talk contains a short introduction to mesonic
 Chiral Perturbation Theory (ChPT). In addition four disparate areas where
some progress has been made in recent years are discussed. These are
the last fit of the order $p^4$ low-energy-constants $L_i^r$ to data,
hard pion ChPT, the recent two-loop work on EFT for QCD like theories
and the high order leading logarithm calculations in the massive $O(N)$
nonlinear sigma model.
}

\FullConference{The 7th International Workshop on Chiral Dynamics,\\
		August 6 -10, 2012\\
		Jefferson Lab, Newport News, Virginia, USA}

\renewcommand{\logo}
{\raisebox{-5mm}
{\parbox{\textwidth}{\begin{flushright}
LU TP 13-03\\
Januray 2013
\end{flushright}
\includegraphics{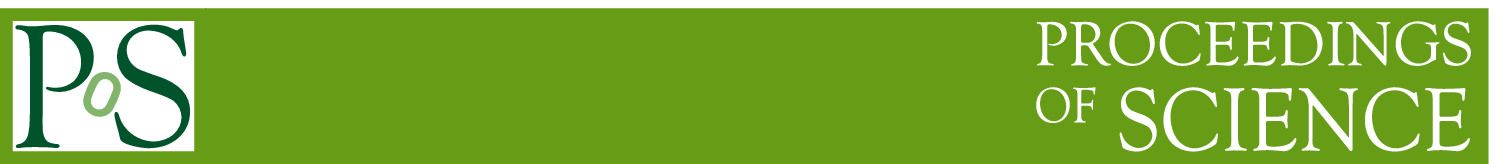}}
}}
\FullConference{{\rm To be published in the proceedings of:\\}
The 7th International Workshop on Chiral Dynamics,\\
		August 6 -10, 2012\\
		Jefferson Lab, Newport News, Virginia, USA}
\begin{document}

\section{Introduction}

This talk is dedicated to my friend and collaborator Ximo Prades who died
since the previous Chiral Dynamics in Berne in 2009. For those of you who
like to know more I recommend looking up the slides from my talk at the
symposium in Granada in his memory \cite{Ximo}. We have been very frequent
collaborators for a long time.

I will not attempt to give an overview of Chiral Perturbation Theory (ChPT)
for mesons in this talk. Even with the restrictions to mesons the subject is
very large. I have given several review talks earlier
\cite{talk1,talk2} as well as written a review article of mesonic ChPT
at two-loop order \cite{review}. References to other reviews and lectures
can be found on my ChPT webpage \cite{webpage}, the remainder there
has admittedly a strong bias towards my own work.

I will give a short introduction to ChPT where I emphasize the major ideas
underlying the method. Afterwards I will give an overview of the work that
has been happening in Lund in this area in the last few years. In addition there
are many more talks at this conference related to mesonic ChPT. As far as the
plenary talks are concerned these are the experimental tests from NA48 \cite{Bizzetti} and KLOE \cite{Bossi}, the connection with dispersion relations
\cite{Lanz,Caprini} and the connection with lattice QCD
\cite{Lellouch,Sachrajda,Izubuchi}. Most of the talks in the working group on
Goldstone Bosons also fall under the topic of this talk.

The remainder is first a short introduction to ChPT, then an overview
of the latest fit of the low-energy-constants,
a few remarks about hard-pion-ChPT as well as some comments about
applications of ChPT beyond QCD and some
leading logarithm calculations to high orders. The treatment is extremely
cursory for all cases.

\section{Chiral Perturbation Theory}

ChPT in its modern form was introduced by Weinberg \cite{Weinberg0}, and
Gasser and Leutwyler \cite{GL1,GL2}. The best way to characterize
ChPT is:\\[1mm]
\centerline{\framebox{\parbox{10cm}{
\centerline{
Exploring the consequences of
the chiral symmetry of QCD}
\centerline{and its spontaneous breaking}
\centerline{using effective field theory techniques}}}}\\[1mm]
A good reference that shows in detail all the underlying assumptions is
\cite{Leutwyler1}.

For an effective field theory, one needs to indicate three things:
the relevant degrees of freedom, a power-counting principle to have predictivity
and the associated range of validity.
For ChPT these are
\begin{description}
\setlength{\itemsep}{0cm}
\setlength{\parsep}{0cm}
\item[Degrees of freedom:] The Goldstone Bosons from the spontaneous
breaking of the chiral symmetry present in QCD in the massless limit.
\item[Power-counting:] Dimensional counting in momenta and masses where
meson masses and momenta are counted the same in the standard counting.
\item[Range of validity:] The breakdown scale is when effects that
are not included become important. For mesonic ChPT these
are clearly the meson resonances in the relevant channels.
$M_\rho$ is thus clearly the end of the range.
\end{description}
Let me show these now in a little more detail. Quantum Chromodynamics (QCD)
for $n_F$ equal mass quarks has an obvious global symmetry under the continuous
interchange of the quarks $SU(n_F)$. Looking at
the purely strong Lagrangian (density)
\begin{equation}
{\cal L}_{QCD} =  \sum_{q=u,d,s}
\left[i \bar q_L D\hskip-1.3ex/\, q_L +i \bar q_R D\hskip-1.3ex/\, q_R
- m_q\left(\bar q_R q_L + \bar q_L q_R \right)
\right],
\end{equation}
we see that in the limit of $m_q=0$ there is actually the larger
$SU(n_F)\times SU(n_F)_R$ global symmetry. Another way to see is that in the
massless case left- and right-handed are no longer related by 
Lorentz-transformations since we cannot go the rest-frame of a massless particle.

\begin{figure}
\begin{minipage}{4.1cm}
\begin{overpic}[width=4cm,unit=0.5pt]{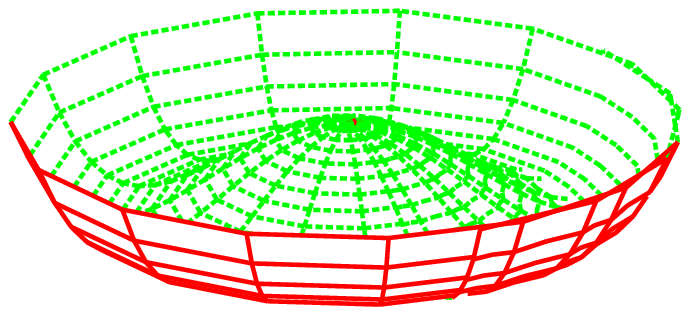}
\SetScale{0.5}
\SetWidth{2.}
\SetColor{Red}
\LongArrow(105,60)(180,20)
\end{overpic}
\caption{\label{fig1} We need to pick a vacuum expectation value indicated
by the arrow. The massless mode corresponds to an excitation along the bottom
of the valley.}
\end{minipage}
\hfill
\begin{minipage}{10cm}
\hspace*{1cm}{rules:}\hspace*{4cm}{ one loop example:}\\[-1cm]
\begin{minipage}{0.3\textwidth}
\unitlength=0.5pt
\begin{picture}(100,100)
\SetScale{0.5}
\SetWidth{1.5}
\Line(0,100)(100,0)
\Line(0,0)(100,100)
\Vertex(50,50){5}
\end{picture}
\hfill\raisebox{25pt}{$p^2$}\\[0.25cm]
\unitlength=0.5pt
\begin{picture}(100,30)
\SetScale{0.5}
\SetWidth{1.5}
\Line(0,15)(100,15)
\end{picture}
\hfill\raisebox{5pt}{$1/p^2$}\\[0.25cm]
$\int d^4p$\hfill$p^4$
\end{minipage}
\hfill
\raisebox{0.75cm}{
\begin{minipage}{0.60\textwidth}
\begin{picture}(100,100)
\SetScale{0.5}
\SetWidth{1.5}
\SetColor{Red}
\Line(0,100)(20,50)
\Line(0,0)(20,50)
\Vertex(20,50){5}
\CArc(50,50)(30,0,180)
\CArc(50,50)(30,180,360)
\Vertex(80,50){5}
\Line(80,50)(100,100)
\Line(80,50)(100,0)
\end{picture}
\hskip-1cm\raisebox{25pt}
{$(p^2)^2\,(1/p^2)^2\,p^4 = p^4$}\\[0.25cm]
\unitlength=0.5pt
\begin{picture}(100,100)
\SetScale{0.5}
\SetWidth{1.5}
\SetColor{Red}
\Line(0,0)(50,40)
\Line(0,50)(50,40)
\CArc(50,70)(30,0,180)
\CArc(50,70)(30,180,360)
\Vertex(50,40){5}
\Line(50,40)(100,50)
\Line(50,40)(100,0)
\end{picture}
~~\raisebox{25pt}
{$(p^2)\,(1/p^2)\,p^4 = p^4$}
\end{minipage}
}
\caption{\label{fig2} An example of the power-counting, on the left the rules
for the lowest order vertex, propagator and loop integration. On the right
showing how it leads to the two one-loop diagrams being $p^4$ while the
tree level lowest-order vertex is $p^2$.} 
\end{minipage}
\end{figure}

The global chiral symmetry $SU(3)_L\times SU(3)_R$ in QCD is spontaneously
broken to the vector subgroup $SU(3)_V$ by the vacuum expectation value
$\langle \bar q q\rangle = \langle \bar q_L q_R+\bar q_R q_L\rangle
\ne 0$. Since in this process 8 generators are broken we get 8 Goldstone
Bosons and their interactions vanish at zero momentum. This is shown pictorially
for one broken generator in Fig.~\ref{fig1}.
The fact that the interaction vanishes at zero momentum allows us to
introduce a proper power-counting along the lines discussed in \cite{Weinberg0}.
This is shown for the example of $\pi\pi$-scattering at one loop in Fig.~\ref{fig2}.

The basic principle just described has been used in very many circumstances.
One needs to decide which chiral perturbation theory
is appropriate for the problem at hand. Some examples are
\begin{itemize}
\setlength{\topsep}{0cm}
\setlength{\itemsep}{0cm}
\setlength{\parsep}{0cm}
\item Which chiral symmetry:
  $SU(N_f)_L\times SU(N_f)_R$, for $N_f=2,3,\ldots$ and
  extensions to (partially) quenched
\item Or beyond QCD
\item Space-time symmetry:
  Continuum or broken on the lattice:
    Wilson, staggered, mixed action
\item Volume: Infinite, finite in space, finite T
\item Which interactions to include beyond the strong one
\item Which particles included as non Goldstone Bosons
\item My general belief: if it involves soft pions (or soft $K,\eta$) some
version of ChPT exists.
\end{itemize}
An important technical step is the inclusion of external fields or sources.
This allows to make the chiral symmetry local\footnote{It is not a gauge symmetry since no kinetic terms for the external fields are included.} \cite{GL1,GL2}.
The manifold at the bottom of the potential in a symmetry breakdown
$SU(n_F)_L\times SU(n_F)\times SU(n_F)_V$ is also an $SU(n_F)$ manifold.
We thus parametrize this by an $n_F\times n_F$ matrix which for $n_F=3$
is
\begin{equation}
U(\phi) = \exp(i \sqrt{2} \Phi/F_0),\quad\mathrm{with}\quad
\Phi (x) = \, \left( \begin{array}{ccc}
\displaystyle\frac{ \pi^0}{ \sqrt 2} \, + \, \frac{ \eta_8}{ \sqrt 6}
 & \pi^+ & K^+ \\
\pi^- &\displaystyle - \frac{\pi^0}{\sqrt 2} \, + \, \frac{ \eta_8}
{\sqrt 6}    & K^0 \\
K^- & \bar K^0 &\displaystyle - \frac{ 2 \, \eta_8}{\sqrt 6}
\end{array}  \right) .
\end{equation}
The lowest-order (LO) Lagrangian is
\begin{equation}
{\cal L}_2 = \frac{F_0^2}{4} \{\langle D_\mu U^\dagger D^\mu U \rangle 
+\langle \chi^\dagger U+\chi U^\dagger \rangle \}\, ,
\end{equation}
with the covariant derivative
$
D_\mu U = \partial_\mu U -i r_\mu U + i U l_\mu \,,
$
and the left and right external currents/fields/sources: $
r(l)_\mu = v_\mu +(-) a_\mu$. 
The scalar and pseudo-scalar external densities are include via $
\chi = 2 B_0 (s+ip).
$ The latter allow the inclusion of quark masses via 
the scalar density: $s= {\cal M} + \cdots$.
The notation implies a flavour trace $\langle A \rangle = Tr_F\left(A\right)$.
The next-to-leading-order (NLO) Lagrangian was worked out by
Gasser and Leutwyler \cite{GL1,GL2} and has 10+2 terms. The
next-to-next-to-leading-order (NNLO) Lagrangian is also known
\cite{BCE1} as well as its infinities \cite{BCE2}. The number of terms is
summarized in Tab.~\ref{tab1}.
\begin{table}
\centerline{\begin{tabular}{|c|cc|cc|cc|}
\hline
      & \multicolumn{2}{c|}{2 flavour} & \multicolumn{2}{c|}{3 flavour} &
\multicolumn{2}{c|}{3+3 PQChPT}\\
\hline
$p^2$ & $F,B$ & 2 & $F_0,B_0$ & 2 &  $F_0,B_0$ &  2 \\
$p^4$ & $l_i^r,h_i^r$ & 7+3 & $L_i^r,H_i^r$ & 10+2 & 
      $\hat L_i^r,\hat H_i^r$ &  11+2 \\
$p^6$ & $c_i^r$ & 52+4 & $C_i^r$ & 90+4 &  $K_i^r$ &
       112+3\\
\hline
\end{tabular}}
\caption{\label{tab1} The number of terms at LO \cite{Weinberg1},
NLO \cite{GL1,GL2}and NNLO \cite{BCE1} for
standard mesonic ChPT.}
\end{table}
 All the cases listed above have extra terms in the Lagrangian with
the exception of finite volume and temperature and in most cases the
Lagrangian is known to NLO. The constants in the Lagrangian are often referred
to as low-energy-constants (LECs).

So, what does ChPT really predict given the large number of free constants.
It relates processes with different numbers of pseudo-scalars and includes
isospin and the eightfold way ($SU(3)_V$). The chiral logarithms can be
seen in the two-flavour ChPT NLO expression for the pion mass\cite{GL1}
\begin{equation}
m_\pi^2 = 2 B \hat m  + \left(\frac{2 B \hat m}{F}\right)^2
\left[ \frac{1}{32\pi^2}{\log\frac{\left(2 B \hat m\right)}{\mu^2}} + 2 l_3^r(\mu)\right] +\cdots
\end{equation}
with $M^2 = 2 B \hat m$ and remember that 
$B\ne B_0$, $F\ne F_0$, the constants in two and three flavour ChPT
are not necessarily the same. The chiral logarithm is the $\log(2 B\hat m)$
term. The infinities are treated by the relation between the bare and
renormalized couplings and the renormalized couplings do depend on
the renormalization scale $\mu$ and the scheme used.

In two-flavour ChPT \cite{GL1} one uses
conventionally the quantities
$\bar l_i = \frac{32\pi^2}{\gamma_i}\, l_i^r(\mu)-\log\frac{M_\pi^2}{\mu^2}\,.
$ which are independent of the scale $\mu$.
For 3 and more flavours, some of the $\gamma_i$ are zero
and one uses the renormalized constants $L_i^r({\mu})$ directly.
The result is in principle independent of $\mu$ but when estimating some of the
constants some choices might be better than others.
The most standard value is $\mu = m_\rho=0.77$~GeV since if we choose
$\mu\approx m_\pi$, $m_K$ the chiral logarithms vanish and if we pick
a larger scale of about 
1 GeV then $L_5^r(\mu)\approx0$ and arguments using a large number of colours,
$N_c\to\infty$ are strongly violated.

A question which often comes up is in which quantities to expand, in
particular whether to use lowest-order or physical quantities. I would like
to stress here that the expansion is in momenta and masses but that it is
\emph{not} unique. There simply is no best way to do the expansion
since relations between the masses, e.g. the Gell-Mann--Okubo relation, exist.
But even more, often there are relations between the kinematical quantities
and masses. An example is $\pi K$ scattering with
$s+t+u=m_\pi^2+m_K^2$, or one can even use the scattering angle as a
kinematical quantity. A related note is that there can be remaining
$\mu$-dependence in a calculation where the $\mu$-dependence is then higher
order. A naive example was discussed in the talk and can be found
in \cite{latticetalk} showing that the apparent convergence of the chiral
series can differ very much depending on what quantities one expands in.
In general, I prefer using the physical masses since thresholds are correct
and the chiral logarithms do come from physical particles propagating.
However, sometimes there are simply too many choices of physical masses
possible, especially in partially quenched and staggered ChPT, and for
the sake of simplicity it might be easier to keep lowest-order masses
everywhere.

\section{Determining low-energy-constants in the continuum}

Lattice QCD has now reached the stage where they can start determining
the ChPT LECs. A review is the FLAG collaboration \cite{FLAG} and
talks at this conference that had results on LECS are \cite{Lellouch,Stolz,
Deuzeman,Bernard}.

For the two-flavour constants the status has not really changed much since
the previous chiral dynamics.
The constants $\bar l_1$ to $\bar l_4$
are determined from ChPT at order $p^6$ and the Roy equation analysis
in $\pi\pi$ and $F_S$ \cite{CGL}. A related talk is \cite{Rios}.
$\bar l_5$
and $\bar l_6$ come from $F_V$ and $\pi\to\ell\nu\gamma$
\cite{BCT,BT}
and from $\Pi_V-\Pi_A$ \cite{GPP}.
Some related work using similar sum rules is \cite{Dominguez}.
In conclusion:
\begin{align}
\label{valueli}
\bar l_1&=-0.4\pm 0.6\,,&
\bar l_2&= 4.3\pm0.1\,,&
\bar l_3&=2.9\pm2.4\,,
\nonumber\\
\bar l_4&=4.4\pm0.2\,,&
\bar l_5 &= 12.24\pm0.21\,,&
\bar l_6-\bar l_5& = 3.0\pm0.3\,,
\nonumber\\
\bar l_6 &= 16.0\pm0.5\pm0.7\,.
\nonumber
\end{align}
and $l_7\sim 5\cdot 10^{-3}$
from $\pi^0$-$\eta$ mixing \cite{GL1}.
In the two-flavour case the contribution from the order $p^6$ LECs was small
so the uncertainty on estimating those is not very important.

\begin{figure}
\centerline{\includegraphics[angle=270,width=0.9\textwidth]{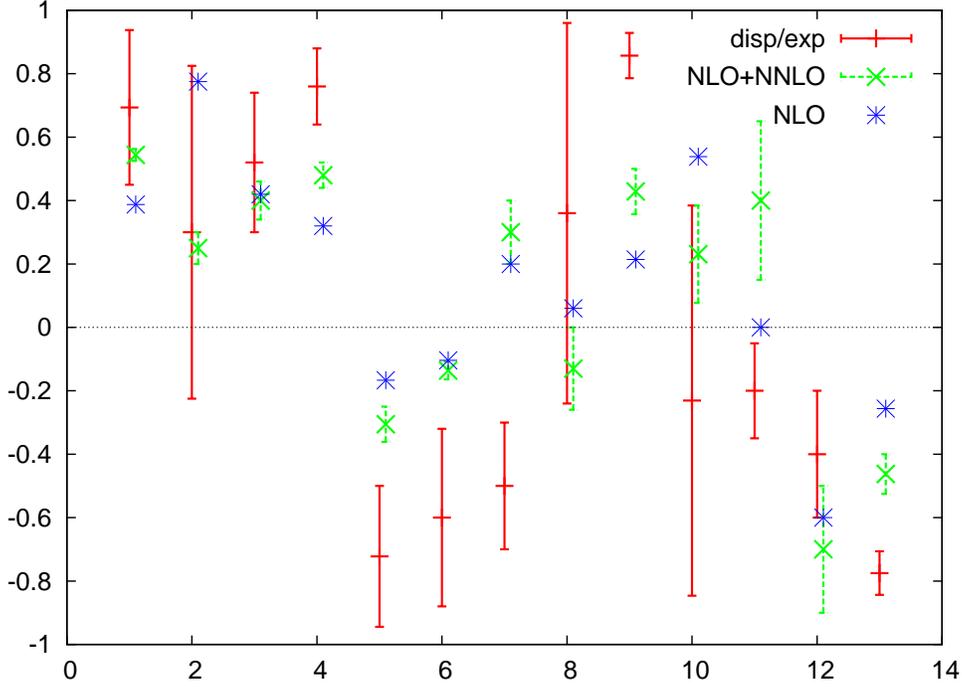}}
\caption{\label{fig3} The quality of the relations found in \cite{relations}.
Shown are the NLO and NNLO predictions for combinations of observables with vanishes NNLO LEC contributions as well as the experimental/dispersive results for the same combinations.}
\end{figure}

The uncertainty on estimates of the order $p^6$ LECs, the $C_i^r$, in the
three flavour case is much more important because $m_K^2,m_\eta^2\gg m_\pi^2$.
The larger $C_i^r$ dependent contributions make their estimates much more
relevant. Typically, if the term multiplying a particular $C_i^r$ leads
to a kinematical dependence it can be measured and the estimates tested.
But if it leads only to a higher order quark-mass dependence they lead
to experiment in a 100\% correlated way with the order $p^4$ LECs, the $L_i^r$.
Here we really need the lattice, typically the latter type of $C_i^r$
need scalars to estimate their values from resonance exchance and are thus 
intrinsically less well estimated. In addition, often the large $N_c$ suppressed
terms in the order $p^6$ lagrangian contribute with large coefficients as well
and all estimates so far of the LECs rely on large $N_c$ to a large extent.
One big question is thus whether we are really testing ChPT or simply
doing a phenomenological fit. 
In order to answer this question a systematic
search for relations that are independent of the $C_i^r$ was done
\cite{relations}. We found there 35 relations between 76 observables
and of these 13 had sufficient data to study their success.
There were 7 relations in $\pi\pi$ scattering which worked about the same for
two and three flavour ChPT. The two relations involving $a_3^-$ did not
work well in either case. Of the 5 extra relations involving
$\pi\pi$ and $\pi K$ scattering three worked well, one of the bad ones involved
$a_3^-$ and the last one had very large NNLO corrections. The final relation
involves $K_{\ell4}$ and did not work well probably because ChPT fails to
describe the quadratic slope in the $F$ form-factor \cite{Kl4,Stoffer}.
The quality of the relations is shown in Fig.~\ref{fig3}. The conclusion is
that three flavour ChPT ``sort of'' works taking into account that the
relations involve very large correlations and that thus the experimental or dispersive input errors might be underestimated.

The previous major fit to determine the $L_i^r$ using NNLO formulas
dates back to 2001 \cite{oldfit}. Given that many more calculations
in three flavour ChPT at NNLO have been performed since and that
the experimental input used then has also changed, an update has become
necessary. This was done in \cite{newfit} and the main results are summarized in Tab.~\ref{tab2}.
\begin{table}
\centerline{\begin{tabular}{|l|c c c c |c|}
\hline
             &fit 10 iso    & NA48/2   & $F_K/F_\pi$ & All $\star$ &  All  \\
\hline
      & old data      &          &            & $\pi\pi$, $\pi K$, $\langle r_S^2\rangle$ & $m_s/\hat m$\\[2mm]
$10^3 L_1^r$ &$0.39\pm0.12$ &\framebox{$0.88$}    & $0.87$     &$0.89$  &$0.88\pm0.09$ \\
$10^3 L_2^r$ &$0.73\pm0.12$ & $0.79$    & $0.80$     &$0.63$  &$0.61\pm0.20$ \\
$10^3 L_3^r$ &$-2.34\pm0.37$& $-3.11$   &$-3.09$     &$-3.06$ &$-3.04\pm0.43$\\
$10^3 L_4^r$ &$\equiv0$     & $\equiv0$ & $\equiv0$  &\framebox{$0.60$} &$0.75\pm0.75$ \\
$10^3 L_5^r$ &$0.97\pm0.11$ & $0.91$     & \framebox{$0.73$}   &\framebox{$0.58$}  &$0.58\pm0.13$ \\
$10^3 L_6^r$ &$\equiv 0$    & $\equiv 0$& $\equiv 0$ & $0.08$ &\framebox{$0.29\pm0.85$}   \\
$10^3 L_7^r$ &$-0.30\pm0.15$& $-0.30$    &$-0.26$    &$-0.22$ &$-0.11\pm0.15$  \\
$10^3 L_8^r$ &$0.60\pm0.20$ & $0.59$     & $0.49$    &$0.40$  &$0.18\pm0.18$  \\ \hline
$\chi^2$  & $0.26$  &$0.01$  &$0.01$   &$1.20$  &$1.28$ \\
dof  & 1 & 1 & 1 & 4 & 4 \\
\hline
\end{tabular}}
\caption{\label{tab2} The main new fit of the $L_i^r$ of \cite{newfit}.
The various columns show the changes w.r.t. the old fit of \cite{oldfit}.}
\end{table}
The old programs in the isospin conserving versions were rewritten
in {C++} for this purpose.
The first column uses the old input values and reproduces the old
fit \cite{oldfit} which was done isospin violation included. The next columns
give the results from including the NA48/2 data on the form factors
\cite{NA482},
the new value of $F_K/F_\pi$, the inclusion of $\pi\pi$, $\pi K$ scattering
and the scalar radius and finally the new value of $m_s/\hat m$. The LEC
that changed most at each step is put in a box.
In the final result one note significant
values for the $N_c$-suppressed constants $L_4^r,L_6^r$, albeit with large errors
and the fact the the large $N_c$-relation $2L_1^r=L_2^r$ no longer holds well.

A large number of variations on the fit were tried in \cite{newfit} by letting
$\mu$ free and varying the input used for the $C_i^r$. All of these gave
similar values for the $L_i^r$. However, a problem occurred when including
the relation between the 2 and 3-flavour LECs of \cite{GHIS}. In order for
getting this relation to work well we needed nonzero values for at least some
$N_c$-suppressed $C_i^r$. In \cite{newfit} a large effort was done to find
reasonable looking $C_i^r$ that allowed to get good fits to all of the inputs.
Many were found but there is no good ground to prefer any of these.
More work especially in trying to include lattice results is definitely needed.
For now, fit All of Tab.~\ref{tab2} and \cite{newfit} should be regarded
as the standard values for the $L_i^r$.

\section{Hard pion ChPT}

The usual formulation of the powercounting in mesonic ChPT \cite{Weinberg0}
assumes that all the momenta in all diagrams are soft and this allows the
powercounting to work as simple dimensional counting. In baryon and heavy
meson ChPT one takes a step further. There is a line with a heavy mass
running through all diagrams but in a way all spatial momenta can still be
considered small. In vector meson ChPT it has been argued that it is possible
to also include diagrams where lines take a hard momentum, but not a hard mass
\cite{HP0}. This type of arguments was used by Flynn and Sachrajda \cite{HP1}
to obtain results for $K_{\ell3}$ in the heavy Kaon limit also away from the
end-point. These arguments, as described below were generalized and applied
to a larger number of processes in \cite{HP2,HP3,HP4,HP5}. Some doubt on
the simple arguments has been presented in \cite{Procura,HP6}.

The underlying idea behind hard pion ChPT is that nonanalyticities in the
the light masses, e.g. pion, come from soft lines in the diagrams.
The couplings of soft particles and in particular soft pions are constrained by
current algebra via
\begin{equation}
\lim_{q\to 0}\langle\pi^k(q)\alpha|O|\beta\rangle
 = -\frac{i}{F_\pi}\langle\alpha|\left[Q_5^k,O\right]|\beta\rangle\,,
\end{equation}
and nothing prevents hard pions to be present in the states $|\alpha\rangle$
or $|\beta\rangle$. One thus expects that by heavily using current algebra
it should be possible to obtain chiral logarithms for almost any process.
 one can always expand in
soft momenta over hard momenta/large masses in a way which is analogous to
the treatment of infrared divergences in QED. The general argument was
already described in my previous chiral dynamics talk \cite{talk2} and
can be found in \cite{HP2,HP3,HP4} as well. It roughly goes as follows:
Take any diagram with a particular external and internal momentum configuration.
Identify the soft lines and cut them.
The resulting part is analytic in the soft stuff and
should thus be describable by an effective Lagrangian with
coupling constants dependent on the external given
momenta (Weinberg's folklore theorem \cite{Weinberg0})
Lagrangian in hadron fields with all orders of derivatives.
This effective Lagrangian should be thought of as a Lagrangian in hadron fields
but all possible orders of the momenta included: possibly
an infinite number of terms
If symmetries are present, the Lagrangian should respect them.
The problem is that the simple power-counting is gone
In some cases we can argue that up to a certain order in
the expansion in light masses, not momenta, matrix
elements of higher order operators are reducible to those
of lowest order, first done in \cite{HP1} for $K_{\ell3}$ and later generalized.
The Lagrangian should be complete in the neighbourhood of the
original process and loop diagrams with this effective Lagrangian should
reproduce the non-analyticities in the light masses.
The latter is the crucial part of the argument.

\begin{figure}
\begin{minipage}{0.48\textwidth}
\includegraphics[angle=270, width=0.99\textwidth]{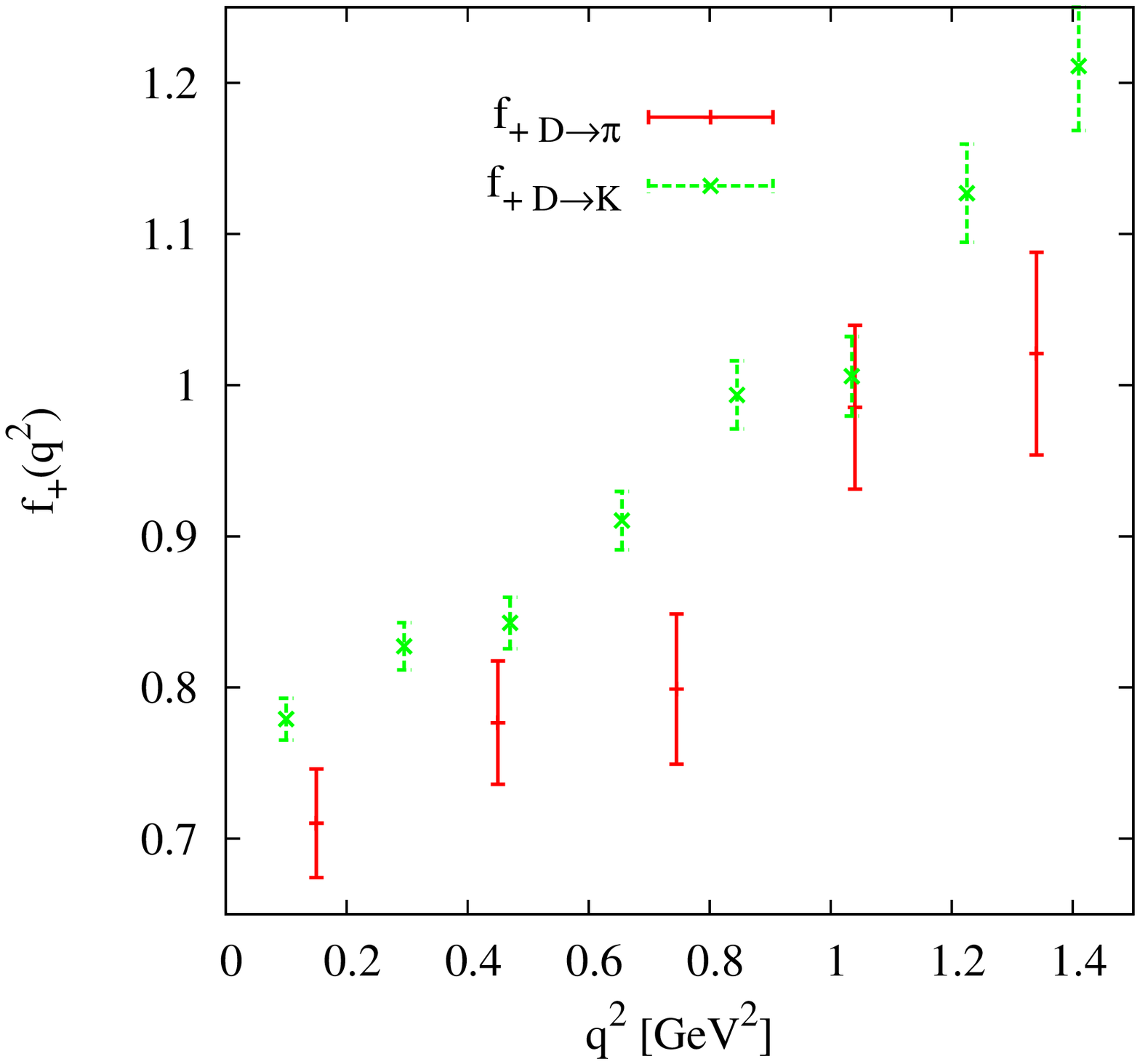}
\end{minipage}
\begin{minipage}{0.48\textwidth}
\includegraphics[angle=270, width=0.99\textwidth]{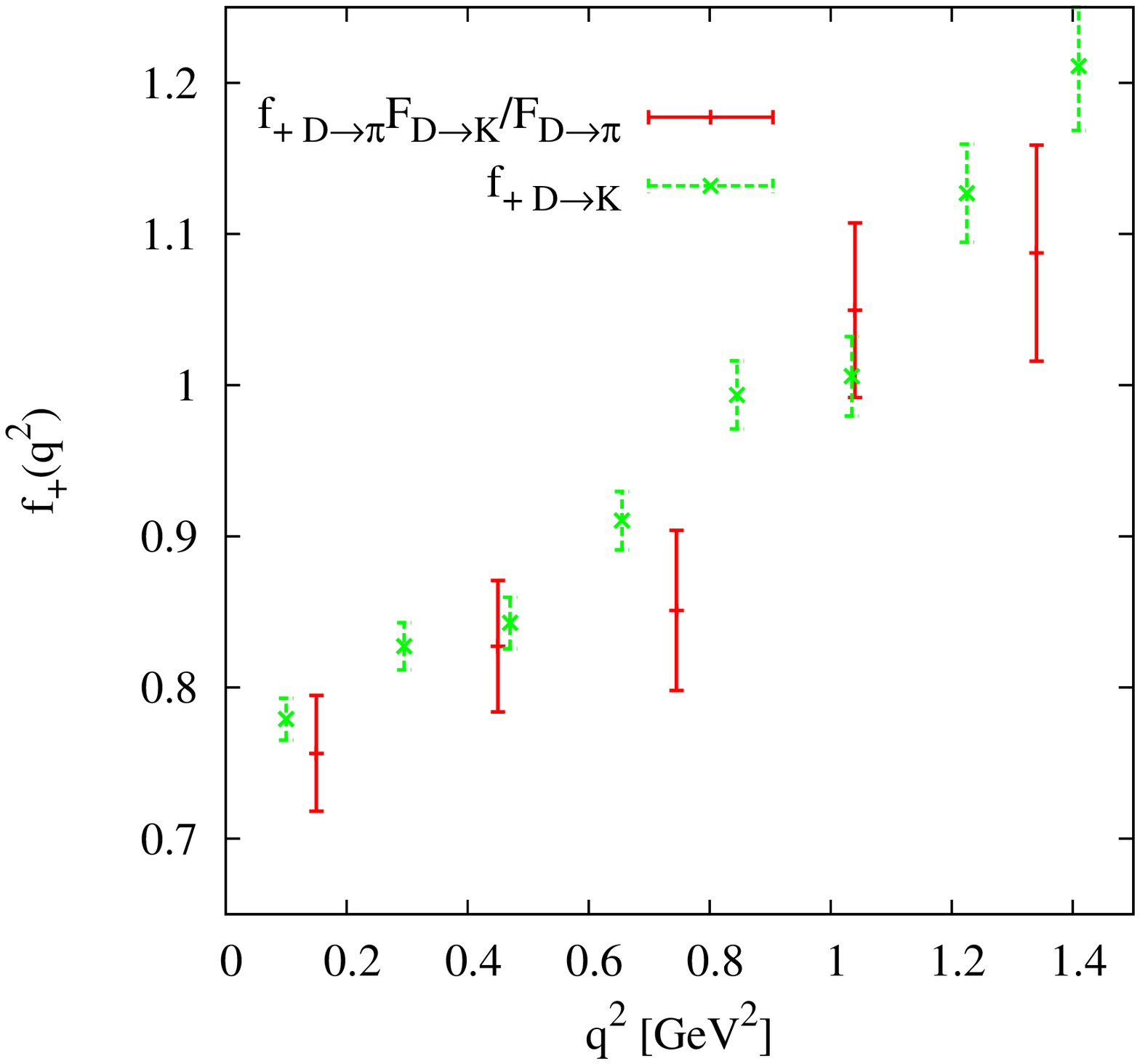}
\end{minipage}
\caption{\label{fig4} The left plot shows the form-factor $f_+$ in $D\to\pi e\nu$
and $D\to K e\nu$ decays from CLEO \cite{CLEO}. In the right hand side
the $D\to K e\nu$ has been corrected with the chiral logarithms via
$f_{+D\to\pi}=f_{+D\to K} F_{D\to\pi}/F_{D\to K}$ \cite{HP4}.}
\end{figure}

A check at two-loop order has been performed for the pion vector and scalar
form-factor in \cite{HP3,HP4} and was found to work well. Applications to the
$B,D\to\pi,K,\eta$ vector form-factors can be found in \cite{HP4} and to
some charmonium decays in \cite{HP5}. In Fig.~\ref{fig4} I show how the
inclusion of the chiral logarithms improves the relation between the
$D\to\pi$ and $D\to K$ form-factors \cite{HP4} using the CLEO data \cite{CLEO}.

Recently \cite{Procura,HP6}, the test done at two loops was extended to
higher orders. What was found there was that the hard pion ChPT prediction
held to all orders for the elastic intermediate state but failed in a
three loop calculation of the inelastic four-particle-cut part.
For details I refer to \cite{Procura,HP6}.
The arguments for the general method are the same as for
IR divergences, SCET,\ldots, so I do believe these to be valid.
Something like hard pion ChPT should exist. 
The arguments for the proportionality to the lowest order
are much weaker and
assume that each soft propagator has a free momentum. The calculation of
\cite{HP6} was done using dispersive methods and a full calculation
at 3 loops will be very difficult but would allow a better study of why the
arguments failed. The ultimate would of course be to find
a proper power-counting under the given assumptions.

\section{Beyond QCD}

The methods of effective field theory and in particular ChPT can also be
used for extensions. A simple extension is to take QCD with $N$ light
flavours but one can also envisage using $N$ fermions in different
representations of the gauge group. The representations can be complex, real or
pseudo-real so we have in general three generic cases to study for
the spontaneous breaking of the global symmetries \cite{Peskin,Preskill}
\begin{equation}
SU(N)\times SU(N)\to SU(N),\quad
SU(2N)\to SO(2N),\quad
SU(2N)\to Sp(2N)\,.
\end{equation}
The cases correspond to a complex, real and pseudo-real representation for
the $N$ fermions. The global symmetry group in the latter two cases
is $SU(2N)$ since both fermion and antifermion are in the same representation
of the gauge group.
Many one-loop results existed especially for the first case,
the equal mass case has been pushed to two-loop order in
\cite{BL1,BL2,BL3}. The main observation \cite{BL1} is that the whole
machinery developed for ChPT can be brought over with a few simple
modifications to the three cases given above. This allowed us to perform
the calculations for the mass and decay constants \cite{BL1},
meson-meson scattering \cite{BL2} and the electroweak precision
parameters \cite{BL3}. The main idea was that lattice calculation can use
our formulas to extrapolate to the massless case, see e.g. \cite{Buchoff}.

The main trick involved is that in all cases mesons can be described by a
unitary matrix $U=\exp(i\phi^a X^a/(\sqrt{2}F))$ with the $X^a$
a different set of generators for the three cases. All flavour sums in
the equal mass case can be done using the relations
\newcommand{\trfa}[1]{\langle #1\rangle}
\begin{align}
\trfa{X^a A X^a B}&= \trfa{A}\trfa{B}
 -\frac{1}{N_F}\trfa{AB}\,,&
\trfa{ X^a A}\trfa{X^a B}&=
 \trfa{AB}
 -\frac{1}{N_F}\trfa{A}\trfa{B}\,.
\nonumber\\
\trfa{X^a A X^a B}&= \frac{\trfa{A}\trfa{B}}{2}
 +\frac{1}{2}\trfa{AJ_S B^T J_S}-\frac{\trfa{AB}}{2N_F}\,,&
\trfa{ X^a A}\trfa{X^a B}&=
 \frac{1}{2}\trfa{AB}+\frac{1}{2}\trfa{AJ_S B^T J_S}
 -\frac{\trfa{A}\trfa{B}}{2N_F}\,.
\nonumber\\
\trfa{X^a A X^a B}&= \frac{\trfa{A}\trfa{B}}{2}
+\frac{1}{2}\trfa{AJ_A B^T J_A} -\frac{\trfa{AB}}{2N_F}\,,&
\trfa{ X^a A}\trfa{X^a B}&=
 \frac{1}{2}\trfa{AB}-\frac{1}{2}\trfa{AJ_A B^T J_A}
 -\frac{\trfa{A}\trfa{B}}{2N_F}.
\end{align}
The lines are for the complex, real and pseudo-real case and
$J_S=
\left(\begin{array}{cc} 0 & I\\
I & 0\end{array}\right)$, $J_A=
\left(\begin{array}{cc} 0 & -I\\
I & 0\end{array}\right)$.

As an example I quote the results for the vacuum expectation value $\langle\bar q q\rangle$ for all three cases.
\begin{align}
\langle \overline q q\rangle &=
\langle \overline q q\rangle_\mathrm{LO}+
\langle \overline q q\rangle_\mathrm{NLO}+
\langle \overline q q\rangle_\mathrm{NNLO}\,.
\nonumber\\
\langle \overline q q\rangle_\mathrm{LO} &\equiv \sum_{i=1,N_F}
\langle \overline q_{Ri}q_{Li}+\overline q_{Li}q_{Ri}\rangle_\mathrm{LO}
= -N_F B_0 F^2
\nonumber\\
\langle \overline q q\rangle_\mathrm{NLO}
 &= \langle \overline q q\rangle_\mathrm{LO}
\left(a_V \frac{\overline A(M^2)}{F^2}+b_V\frac{M^2}{F^2}\right)\,,
\nonumber\\
\langle \overline q q\rangle_\mathrm{NNLO}
 &= \langle \overline q q\rangle_\mathrm{LO}
\Bigg(c_V \frac{\overline A(M^2)^2}{F^4}
     +\frac{M^2\overline A(M^2)}{F^4}\left(d_V+\frac{e_V}{16\pi^2}\right)
  +\frac{M^4}{F^4}\left(f_V+\frac{g_V}{16\pi^2}\right)\Bigg)\,.
\label{xvev}
\end{align}
I used here 
$M^2 = 2 B_0\hat m$ and $
\overline A(M^2) = -\frac{M^2}{16\pi^2}\log\frac{M^2}{\mu^2}\,.
$ The coefficients appearing are given in Tab.~\ref{tab3}.
Note the similarity between the different results and the existence of
a number of large $n$ relations between the various cases \cite{BL1}.
\begin{table}
\begin{tabular}{|c|c|c|c|}
\hline
 & QCD & Adjoint & 2-colour \\
\hline
$a_V$ & $n-\frac{1}{n}$ & $ n+\frac{1}{2}-\frac{1}{2n}$
 & $ n-\frac{1}{2}-\frac{1}{2n}$\\
$b_V$ & $16 n L_6^r+8 L_8^r +4 H_2^r$& $32 n L_6^r+8 L_8^r +4 H_2^r$
 & $32 n L_6^r+8 L_8^r +4 H_2^r$ \\
$c_V$ & $\frac{3}{2}\left(-1+\frac{1}{n^2}\right)$& $\frac{3}{8}\left(-1+\frac{1}{n^2}-\frac{2}{n}+2n\right)$
 & $\frac{3}{8}\left(-1+\frac{1}{n^2}+\frac{2}{n}-2n\right)$ \\
$d_V$ & $-24\left(n^2-1\right)\,
  \left(L_A+\frac{L_B}{n}\right)$& $-12\left(2n^2+n-1\right)
  \left(2L_A+\frac{L_B}{n}\right)$
 &  $-12\left(2n^2-n-1\right)
  \left(2L_A+\frac{L_B}{n}\right)$\\
$e_V$ & $1-\frac{1}{n^2}$&$\frac{1}{4}\left(1-\frac{1}{n^2}+\frac{2}{n}-2n\right)$
& $\frac{1}{4}\left(1-\frac{1}{n^2}-\frac{2}{n}+2n\right)$ \\
$f_V$ & $48\left(K_{25}^r+n K_{26}^r+n^2 K_{27}^r\right)$& $r^r_{VA}$ 
& $r^r_{VT}$ \\
$g_V$ & $8\left(n^2-1\right)\,
\left(L_A+\frac{1}{n}L_B\right)$& $4\left(2n^2+n-1\right)
  \left(2L_A+\frac{1}{n}L_B\right)$
 & $4\left(2n^2-n-1\right)
  \left(2L_A+\frac{1}{n}L_B\right)$ \\
\hline
\end{tabular}
\caption{\label{tab3} The coefficients appearing 
the corrections to the vacuum-expectation-value for the three generic cases
with $n$ flavours. Table from \cite{BL1}.
We have defined the abbreviations 
 $L_A=L_4^r-2L_6^r$ and
$L_B=L_5^r-2L_8^r$.}
\end{table}

\section{Leading logarithms}

The last part on which I want to report is some recent progress
in calculating leading logarithms in effective field theories.
Some of this is also mentioned in the talk by Kampf \cite{Kampf}.
The underlying argument goes something like:
Take a quantity with a single scale: $F(M)$.
The dependence on the scale in field theory is typically
logarithmic, so with $L=\log\left(\mu/M\right)$we get
\begin{equation}
F= F_0 + F_1^1 {L} + F^1_0 + F_2^2{ L^2} + F^2_1 L + F^2_0 
+ F_3^3{ L^3}
 +\cdots
\end{equation}
The leading logarithms are the terms with $F_m^m{ L^m}$.
The $F_m^m$ can be more easily calculated than the
 full result. This follows from the fact that for any physical quantity
$\mu\left( dF/d\mu\right)\equiv 0$ and 
ultraviolet divergences in Quantum Field Theory are always local.
In renormalizable theories this is embedded in the renormalization group but for
effective field theories such as QCD there is no simple recursive argument.
Weinberg already argued that one can get away with only one-loop calculations
to obtain the leading logarithms \cite{Weinberg0}. This was proven in \cite{BC0} and in a somewhat simpler way in \cite{BC1}. The underlying reason is that the cancellation of nonlocal divergences gives a set of consistency relations
between contributions of different loop order as explained in \cite{BCEGS}.
In the massless case \cite{Polyakov1,Polyakov2,Polyakov3} this leads to
an almost analytic expression to very high orders since the diagrams remain
fairly simple to all orders. In the massive case diagrams with any number of
external legs show up but the whole process can be automatized since
in \cite{BC1} it was realized that the Lagrangians at higher order do not need
to be minimal. Obtaining the minimal Lagrangian at each order would have been
essentially impossible. In the massive case all published results relate to
the $O(N)$ model, masses in \cite{BC1}, decay constants and vacuum expectation
values as well as form-factors and meson-meson scattering in \cite{BC2} and
the anomalous sector \cite{BKL}. Extension to $SU(N)\times SU(N)$ has been
done in the massless case \cite{Polyakov3} and is in progress for the massive
case.
I refer to the original papers for more results but a few highlights are
that the large $N$ (number of flavours) limit is not a good approximation
for any of the quantities calculated. The series seem to converge
in the expected regions. None of the leading logarithms calculated seems
to be unusually large. Unfortunately the hope that we might recognize
the general result for arbitrary $N$ proved in vain.

\acknowledgments
I thank the organizers for a very pleasant meeting and my collaborators in the
various parts reported here.
This work is supported in part by the European Community-Research
Integrating Activity ``Study of Strongly Interacting Matter''
(HadronPhysics3, Grant Agreement No. 283286)
and the Swedish Research Council grants 621-2011-5080 and 621-2010-3326.

\end{document}